\documentstyle[aps,preprint]{revtex} \baselineskip 12pt
\begin{document}
\title{ DO QUASI-EXACTLY SOLVABLE SYSTEMS ALWAYS CORRESPOND TO
ORTHOGONAL POLYNOMIALS? 
 }

\vspace{.4in}

\author{Avinash Khare and Bhabani Prasad Mandal}

\address{
Institute of Physics, Sachivalaya Marg,\\ Bhubaneswar-751005, India,\\
Email:  khare, bpm@iop.ren.nic.in}

\vspace{.4in}

\maketitle

\vspace{.4in}

\begin{abstract}
We consider two quasi-exactly solvable problems in one dimension for which
the Schr$\ddot{o}$dinger equation can be converted to Heun's equation. We
show that in neither case the Bender-Dunne polynomials form an orthogonal set.
Using the anti-isopectral transformation we also discover a new
quasi-exactly solvable problem and show that even in this case the
polynomials do not form an orthogonal set.

\end{abstract}

\newpage

Some time ago, in a remarkable paper Bender and Dunne ~\cite{bd} showed that the
eigenfunctions of the Schr$\ddot{o}$dinger equation  for a quasi-exactly
solvable ( QES ) problem is the generating function for a set of orthogonal polynomials
$\{P_n(E)\}$ in the energy variable $E$. It was further shown that these 
polynomials satisfy the three-term recursion relation 
\begin{equation}
P_n(E) = EP_{n-1}(E) +C_nP_{n-2}(E)
\label{r1}
\end{equation}
where $C_n$ is $E$ independent quantity. Using the well known theorem
\cite{the,ext},
`` the {\it necessary and sufficient} condition for a family of polynomials 
$\{P_n\}$ (with degree $P_n= n $) to form an orthogonal polynomial system is 
that $\{P_n\}$ satisfy a three-term recursion relation of the form
\begin{equation}
P_n(E) = \left ( A_n E +B_n \right ) P_{n-1}(E) + C_n P_{n-2}(E) \ \ \ \ \  n\ge 1
\label{r2}
\end{equation}
where the coefficients $A_n , B_n $ and $C_n$ are independent of $E$, 
$A_n \neq 0 , C_1 =0 , C_n \neq 0 $ for $n\ge 1$" , it then followed that
$\{P_n(E)\}$ for this problem forms an orthogonal set of polynomials with 
respect to some weight function, $w(E)$.
Recently several authors have studied  the Bender-Dunne polynomials 
in detail \cite{ext,two,two1,bd1}. In fact it has been claimed that the Bender-Dunne 
construction is quite universal and valid for any quasi-exactly solvable 
model in both one as well as  multi-dimensions. 

The purpose of this note is to critically examine this assertion. In
particular,
 we discuss two QES systems, one of which is in one dimension while the other
one is an effective one dimensional system. We show in both the cases 
that though these problems are QES problems, the corresponding three-term recursion relation 
is not of the type as given in Eq. (\ref{r2}), and hence the corresponding 
polynomials do not form an orthogonal set.
Using the anti-isospectral
transformation $x\rightarrow ix$ introduced recently by Krajewska {\it et
al.}\cite{two1}, we discover a new QES problem and show that even in this case
the polynomials do not form an orthogonal set.

Consider the potential,
\begin{equation}
V(x) = \frac{ \mu^2 \left [ 8\sinh^4 \frac{ \mu x}{2}-4(\frac{ 5}
{\epsilon ^2}-1)\sinh^2 \frac{ \mu x}{2}+2 \left (\frac{ 1}{\epsilon ^4}
-\frac{ 1}{\epsilon ^2}-2 \right ) \right ]}{ 8 \left [1+\frac{ 1}{\epsilon ^2}+\sinh^2 
\frac{ \mu x}{2} \right ]^2}
\label{p1}
\end{equation}
which arises in the context of the stability analysis around the kink
solution for $\phi^6 $-type field theory in $1+1$ dimensions \cite{lee}.
It has already been shown that it is an example of QES system
in which both the ground and the second excited states are exactly known in the 
case $\epsilon ^2 =\frac{ 1}{2}$. Further it has also been shown  that contrary to the usual
belief \cite{book}, this QES problem can not be generated by the quadratic elements
of an enveloping $Sl(2)$ algebra \cite{jat}. A related fact is that the
Schr$\ddot{o}$dinger equation for this case, after some transformations, can be
written in the form of Heun's equation  which has four regular singular
points \cite{lee}. We shall now show that in this case one can obtain 
a three-term recursion relation for the polynomials $\{P_n(E)\}$ which is
different from the one given in Eq. (\ref{r2}) and hence these polynomials 
do {\it not} form an  orthogonal set even though it is a QES system.

Consider the Schr$\ddot{o}$dinger equation ( $\hbar = 2m =1$)
\begin{equation}
\left [ -\frac{ d^2}{dx^2} + V(x) \right ]\psi(x) = E\psi(x)
\label{scw}
\end{equation}
On making use of the transformation
\begin{equation}
y =\frac{  \sinh^2 \frac{ \mu x}{2}}{\left (1+\frac{ 1}{\epsilon ^2 } + \sinh^2 \frac{ \mu x}{2} \right 
) },\ \ \ \psi = (1-y)^s f , \ \ \ s=\left (1-\frac{ E}{\mu^2} \right )^{\frac{ 1}{2}} 
\end{equation}
the Schr$\ddot{o}$dinger equation reduces to the 
Heun's equation,
\begin{equation}
f^{\prime\prime}(y)+ \left [ \frac{ 1}{2y} + \frac{ 1+2s}{y-1} + \frac{ 1}{
2(y+\epsilon ^2)} \right ]f^{\prime}(y) + \frac{ ( \alpha\beta y
-q)}{y(y-1)(y+\epsilon ^2)}f(y)=0
\label{heu}
\end{equation}
Here 
\begin{equation}
\alpha = - \frac{ 5}{2}-s; \ \ \beta = \frac{ 3}{2} -s ; \ \ q= (1-s^2)(1+\epsilon ^2)-
\frac{ 1}{2}s \epsilon ^2- \frac{ 1}{4}(1-2 \epsilon ^2).
\end{equation}
 
On further substituting 
\begin{equation}
t = \left (y+\frac{ 1}{2} \right )^{\frac{ 1}{2}}
\end{equation}
it is easy to show that $f(t)$ satisfies
\begin{eqnarray}
(t^2-\epsilon ^2)(t^2-1-\epsilon ^2)f^{\prime\prime}(t) &&+ \left [
(t^2-1-\epsilon ^2) + 2(1+2s)(t^2-\epsilon ^2)\right
]tf^\prime(t)\hspace{.3in} \nonumber \\
&&+\left [4\alpha \beta t^2 -4(\alpha \beta \epsilon ^2+q) \right ] f(t) =0
\label{heu1}
\end{eqnarray} 
From this equation one can easily obtain the three-term recursion
relation satisfied by the polynomials associated with this QES system.
In particular, on substituting
\begin{equation}
f(t) = \sum_{n=0}^{\infty} \frac{ Q_n(s) t^n}{n!}
\label{13}
\end{equation}
 in Eq. (\ref{heu1}) it is easily shown  that $Q_n$'s
satisfy the three-term recursion relation 
\begin{eqnarray}
\epsilon ^2(\epsilon ^2+1)Q_{n+2}(s) &&- \left [(2 \epsilon ^2+1
)n^2 +(5 \epsilon ^2+2+4s \epsilon ^2)n -4s^2 +2s \epsilon ^2+3 -9 \epsilon ^2\right ]Q_n(s) 
\nonumber\\&&+
 n(n-1) \left [ n^2+n(4s-2)  
+4s^2-4s -15 \right ]Q_{n-2}(s) =0
\end{eqnarray}

Notice that the even and the odd $Q_n$'s are unrelated and we can obtain
two separate recursion relations for them.
In particular it is easily shown that for even and odd cases, the recursion relations 
 respectively are ($m\ge 1$)
\begin{eqnarray}
 \epsilon ^2(\epsilon ^2+1)P_m(s) -&&\left [(8 \epsilon ^2+4)m^2 
+(8s \epsilon ^2-6 \epsilon ^2-4)m -4s^2 -6s \epsilon ^2 +3- 11\epsilon ^2 \right
]P_{m-1}(s) \nonumber \\+&&(m-1)(2m-3) 
\left [8m^2+(16s-24)m+8s^2-24s-14 \right ]P_{m-2}(s) =0
\label{rc1}
\end{eqnarray}
and 
\begin{eqnarray}
\epsilon ^2(\epsilon ^2+1)P_m(s)-\left [(8 \epsilon ^2+4)m^2 +(8s \epsilon ^2
+2 \epsilon ^2)m -4s^2 -2s \epsilon ^2+ 2-12 \epsilon ^2 \right ]P_{m-1}(s)
+\nonumber \\ (m-1)(2m-1) 
\left [8m^2+(16s-16)m+8s^2-16s-24 \right ]P_{m-2}(s) =0
\label{rc2}
\end{eqnarray}
subject to the initial condition, $P_0(s) =1$.

We observe that neither of these three-term recursion relations are of the
form as given by Eq. (\ref{r2}) and hence we conclude that even though it is
a QES problem, the corresponding polynomials $P_m(s)$ are not orthogonal.
Note that the QES solution corresponding to the ground state is obtained
from Eqs. (\ref{13}) and (\ref{rc2}) and in this case $s=1
(i.e.\  E=0)$ and $m=2$( irrespective of the value of $\epsilon $), while the solution
corresponding
to the second excited state is obtained from Eqs. (\ref{13}) and (\ref{rc1}) when $s=\frac{ 1}{2}
(i.e.\  E=\frac{ 3}{4}\mu^2)$ and $m=3$ only if $\epsilon ^2 =\frac{ 1}{2}$. Thus
only for $\epsilon ^2 = \frac{ 1}{2},$ this is a QES problem. 

Why is that in this case the polynomials do not form an orthogonal set?
While we do not have a definite answer, we suspect that only for 
those QES systems for which the symmetry group is $Sl(2) $, the corresponding polynomials form an
orthonormal set. In this context recall that for the above problem, 
the Hamiltonian can not be written in terms of the quadratic generators
of $Sl(2)$ \cite{jat}.  
 As a support to our conjecture, we offer another QES problem where
also the Schr$\ddot{o}$dinger equation can be reduced to Heun's equation with
four regular singular points and hence the corresponding symmetry
group is not $Sl(2)$. In that case also we find the 
three-term recursion relation satisfied by the polynomials and show that it is not 
 of the type as given by  Eq. (\ref{r2}) and hence these polynomials 
also do not form an orthogonal set.

Recently Bhaduri {\it et al.} \cite{bha}  have considered a two body problem characterized
by 
\begin{equation}
H= -\frac{ 1}{2}( \vec{\nabla}_1^2 +\vec{\nabla}_2^2 ) + \frac{ 1}{2}
   \left [ r_1^2+r_2^2 \right ]+ \frac{ g_1}{2}\left ( \frac{
r_1^2+r_2^2}{X^2} \right )
\end{equation}
where $X = x_1y_2-x_2y_1$.
They were able to solve the problem in hyperspherical coordinates, $ (R, \theta, 
\phi, \psi)$. In particular they showed  that the angular equation is
\begin{equation}
\left ( \Lambda^2 + \frac{ 4g_1}{\sin^2(2 \theta ) 
} \right ) \Phi = \beta(\beta+2)\Phi \ \ \ \beta\geq 1
\end{equation}
while the radial equation is of the type of harmonic oscillator in
4-dimensions.
\begin{equation}
\frac{ d^2F}{dR^2} +\frac{ 3}{R}\frac{ dF}{dR} + \left ( 2E-R^2-\frac{
\beta(\beta+2)}{R^2} \right ) F = 0
\end{equation}
Here $\Lambda^2$ is the Laplacian  on the sphere $S^3$.
On making the ansatz,
\begin{equation}
\Phi(\theta, \phi,\psi) = P(x) e^{iq\phi}\ e^{il\psi}
\end{equation}
where $q $ and $l$ are integers and 
\begin{equation}
P(x)= |x|^a(1-x)^b(1+x)^c \Theta^{a,b,c}(x)
\end{equation}
it was shown that $\Theta(x)$ satisfies Heun's equation,
\begin{eqnarray}
(1-x^2)\frac{ d^2\Theta}{dx^2}&&+ 2 \left  [ \frac{ a}{x}-(b-c)-(a+b+c+1)x \right ]
\frac{ d\Theta}{dx}\nonumber \\&&+ \left [\frac{ (\beta+1)^2}{4}- 
(a+b+c+\frac{ 1}{2})^2+ \frac{ 2a(c-b)}{x}\right ] \Theta(x) =0
\label{heu2}
\end{eqnarray}
Here $b= \frac{ |l+q|}{4}, \ \ c= \frac{|l-q|}{4}, \ a(a-1)= g_1 $. 
In order to obtain the recursion relation satisfied by the polynomials
associated with this QES problem we substitute
\begin{equation}
\Theta(x) = \sum_{n=0}^\infty \frac{ P_n(\beta) x^n}{n!}
\end{equation}
in Eq. (\ref{heu2}).
It is easy to show that $P_n(\beta) $ satisfies the three-term recursion
relation 
\begin{eqnarray}
(n+2a-1) P_n -&&2(b-c)(n-1+a)P_{n-1}\nonumber \\  +&& (n-1) \left [
\frac{ (\beta+1)^2}{4}- \left (a+b+c-\frac{ 3}{2} \right )^2 \right ]P_{n-2}
 =0
\end{eqnarray} 
when $n\geq 1$ and $P_0 =1$. We again observe that the recursion relation
is not of the type given by Eq. (\ref{r2})  and hence it follows that the
polynomials $P_n(\beta)$ do not form an orthogonal set. The QES solutions
in this case have already been discussed in Ref. \cite{bha}.

Summarizing, we have seen that there are QES problems in one dimension
and effective one dimension where the symmetry group is not $Sl(2)$, the
Schr$\ddot{o}$dinger equation get converted to Heun's equation, and the
corresponding polynomials do not form an orthogonal set. It will be
interesting to find the symmetry group in these cases.

Before finishing this note we would like to point out a nontrivial
application of the anti-isospectral transformation induced recently
by Krajewska {\it et al.}\cite{two1}. In particular, applying it we
are able to discover a new QES problem in quantum mechanics. Consider the
potential given in Eq. \ref{p1} and apply the transformation $x\rightarrow i\theta
$. We then obtain a new QES problem which has not been discussed in
the literature so far. In particular the ground and second excited state
energies of the new (periodic) potential 
\begin{equation}
V(\theta ) = -\frac{ \mu^2 \left [ 8\sin^4 \frac{ \mu \theta }{2}+4(\frac{ 5}
{\epsilon ^2}-1)\sin^2 \frac{ \mu \theta }{2}+2 \left (\frac{ 1}{\epsilon ^4}
-\frac{ 1}{\epsilon ^2}-2 \right ) \right ]}{ 8 \left [1+\frac{ 1}{\epsilon ^2}-\sin^2 
\frac{ \mu \theta }{2} \right ]^2}
\label{p2}
\end{equation}
are $E_0= - \frac{ 3}{4}\mu^2, E_2 =0$ in case $ \epsilon ^2= \frac{
1}{2}$. The corresponding eigenfunctions  can easily be written
down. Needless to say in this case also one can obtain a three-term
recursion relation and show that it is not of the type as given by Eq. (\ref{r2}). Thus
this provides a third QES problem where the Bender-Dunne polynomials do not
form  an orthogonal set. The details will be given elsewhere \cite{qes2} where
we will also discuss other applications of the anti-isospectral transformation. 


\newpage
\begin{references}
\bibitem{bd} C. M. Bender and G. V. Dunne, J. Math. Phys. {\bf 37}(1996) 6.

\bibitem{the} T.S. Chihara, {\it An Introduction to Orthogonal Polynomials }
, Gordon and Breach, New York, (1975); A. Erdelyi, W. Magnus, F. Oberhettinger
and F. G. Tricomi, {\it Higher Transcendental Functions}, Vol.II, McGraw-Hill
, New York, (1953).
\bibitem{ext}  F. Finkel, A. Gonzalez-Lopez And M. A. Rodriguez,
J. Math. Phys. {\bf 37}(1996) 3954.

\bibitem{two} A. Krajewska, A. Ushveridze and Z. Walczak,
Mod. Phys. Lett. {\bf A 12}(1997) 1131.

\bibitem{two1} A. Krajewska, A. Ushveridze and Z. Walczak,
Mod. Phys. Lett. {\bf A 12}(1997) 1225.

\bibitem{bd1} C. M. Bender, G. V. Dunne and M. Moshe, Phys. Rev. {\bf A 55}(1997) 2625.

\bibitem{lee} N. H. Christ and  T.D. Lee  Phys. Rev. {\bf D 12}(1975)
1606.
\bibitem{book} A. Ushveridze, {\it Quasi-Exactly Solvable Models
in Quantum Mechanics}, Inst. of Physics Publishing, Bristol, (1994). 
\bibitem{jat}D. P. Jatkar, C. Nagaraja Kumar and A. Khare, Phys. Lett.
{\bf A 142}(1989) 200.

\bibitem{bha} R. K. Bhaduri, A. Khare, J. Law, M. V. N. Murthy and D. Sen
, J. Phys. A : Math. Gen. {\bf 30}(1997) 2557.

\bibitem{qes2} A. Khare and B. P. Mandal ( In preparation).
\end{references}
\end{document}